\documentclass{llncs}
\usepackage{graphicx}
\usepackage{epsfig}
\newcommand{\remove}[1]{}

\newtheorem{result}{Result}

\title{Approximation Algorithms for Line Segment Coverage in Wireless Sensor Networks}
\author{Dinesh Dash\inst{1} \and Arijit Bishnu\inst{2} \and Arobinda Gupta\inst{1} \and Subhas C. Nandy\inst{2}}
\institute{Dept. of CSE, Indian Institute of Technology - Kharagpur, India \and 
ACM Unit, Indian Statistical Institute, Kolkata, India}

\begin{document}
\maketitle
\begin{abstract}
The coverage problem in wireless sensor networks deals with the
problem of covering a region or parts of it with sensors. In this
paper, we address the problem of {\em covering a set of line
segments} in sensor networks. A line segment $\ell$ is said to be
\emph{covered} if it intersects the sensing regions of at least
one sensor distributed in that region. We show that the problem of 
finding the minimum number of sensors needed to {\em cover} each member in a  
given set of line segments in a rectangular area is NP-hard. Next, 
we propose a constant factor approximation algorithm for the problem 
of covering a set of axis-parallel line segments. We also show that 
a PTAS exists for this problem. \remove{Though the constant factor of our 
approximation algorithm is high, simulation results show 
that the produced solution is reasonably well in practice.}
\end{abstract}

\section{Introduction}\vspace{-0.1in}
\label{sec:intro} A wireless sensor network (WSN) consists
of a number of tiny devices equipped with sensors to sense one or more
parameters such as temperature, speed etc. Due to the limited battery
power, each sensor node can do a limited amount of computation, and
can communicate with only nearby devices. Each sensor has a sensing
range within which it can sense the parameter, and a communication
range over which it can communicate with other devices. Sensor
networks have been used in different applications such as environment
monitoring, intruder detection, target tracking etc.


The {\em coverage} problem is an important problem in many wireless
sensor network applications. Here, a set of sensors are used for {\em
surveillance} (or {\em monitoring}) of an area. Various definitions of coverage may
be considered depending on the target application. For example, the
{\em $k$-coverage} problem requires that every point in the area be in
the sensing range of at least $k$ sensors \cite{hu2}. In the {\em
target $k$-coverage} problem, a set of points in the plane are marked
as target points; the objective is to place sensors such that every
target point is in the sensing zone of at least $k$ sensors. Other
definitions of coverage include {\em area coverage} \cite{th3}, {\em
barrier coverage} \cite{sa33}, {\em breach} and {\em support} paths
\cite{me5} etc. Several works provide algorithms for achieving various
types of coverages in sensor networks by suitable placement of the
sensors \cite{hu2,sa33,me5,th3,xi60}.

In many applications, it is required that a set of line segments in
a region be covered with sensors. Examples of such applications can 
be monitoring activities in the corridors of a building, or in the 
road networks of a region. A line segment $\ell$ is said to be {\it 
k-covered} if $\ell$ intersects the sensing range of at least $k$ 
sensors. Thus, given a set of line segments in the plane, it may be
required to place a set of sensors to ensure that all the line
segments are $k$-covered. In this paper, we consider the following 
variation of the problem.
\vspace{-0.1in}
\begin{description}
\item[Line-Covering problem:] Given a set $L$ of $n$ arbitrarily oriented
line segments in a bounded rectangular region $R$, find the minimum
number of sensors (with equal sensing range $\rho > 0$) needed, and
their positions such that each line segment in $L$ passes through the 
sensing region of at least one sensor. 
\end{description}

\vspace{-0.1in}
We prove that the decision version of the {\em Line-Covering problem} 
is NP-hard, and present a constant factor approximation algorithm 
for a special case where the line segments in $L$ are all axis-parallel 
(horizontal or vertical). We also show that the {\em Line-Covering problem} 
for axis-parallel line segments admits a PTAS. Note that, there are several 
practical situations, such as surveillance of corridors in a floor, where 
covering axis-parallel line segments with sensors is indeed necessary. 

A variation of line coverage problem, called track coverage problem is 
addressed by Baumgartner et al. \cite{BF}, where the objective is to place 
a set of $n$ sensors in a rectangular region of interest such that a measure 
of the set of tracks detected by at least $k$ sensors is maximized. The 
measure may be the width of track, or the angle of a cone originated from 
one end-point of the track, where the central lines of the tracks are given. 
However, to the best of our knowledge, this problem we are considering here, 
is not addressed in the existing literature. 

\vspace{-0.1in}
\section{Related works}\vspace{-0.1in}
\label{sec:relwork}


Given a deployment of sensors in a bounded region, several algorithms
have been proposed to compute different types of coverage problems. 
Huang and Tseng \cite{hu2} proposed an algorithm for testing whether 
every point in an area is $k$-covered. Xing et al. \cite{xi60} gave 
an algorithm to verify whether an area is connected-covered by a set 
of $k$ sensors. The {\em k-barrier coverage} problem was defined by 
Kumar et al. \cite{sa33}.
They also proposed an efficient algorithm for testing whether a barrier 
is $k$-covered or not. The problems of finding maximal breach and maximal 
support paths were addressed by Megerian et al. \cite{me5}.

The problem of efficient deployment of sensors for efficiently covering 
an area is also studied in the literature. Given a fixed number of sensors
and an area with obstacles, Wu et al. \cite{wu2007} proposed a centralized 
and deterministic sensor deployment strategy in the obstacle free regions 
for maximizing the area covered by the deployed sensors. Agnetis et al. 
\cite{grandeb2009} addressed the problem of deploying sensors with minimum 
cost under a defined cost model for full surveillance, where every point 
on each line segment is covered by at least one sensor. They provided a 
polynomial time algorithm for the case where all sensors have the same 
sensing range. They also provided a branch-and-bound based heuristic for 
some special cases where the sensing ranges of the sensors are different.
Clouqueur et al. \cite{clouqueur2002} presented a deployment strategy
to find a minimum {\em exposure} path for a moving target with minimum
deployment cost, where each sensor has a deployment cost which depends 
on its range. The \emph{exposure} of a path with respect to a target 
through the sensor field is measured in terms of the probability that 
the target will be detected by some sensor along that path. Bai et al. 
\cite{bai2006} proposed an optimal deployment strategy of sensor nodes 
such that these can cover the entire region as well as the communication 
network becomes biconnected.

A sizeable literature exists on maintaining different types
of coverage in wireless sensor networks by moving one or more sensors
after the initial deployment. After an initial random deployment, here
the objective is to maintain the coverage by moving minimum number
of sensors \cite{za25}. Sometimes sensor(s) may need to be moved
due to the failure of other sensors. Sekhar et al. \cite{ar116}
proposed a dynamic coverage maintenance scheme. In their work, if a coverage
hole is created due to the failure of a sensor, only the neighbors of
the dead sensor are migrated to cover that hole with minimum total
energy consumption. There are several other situations where the
sensors may move \cite{139ch,wa133,33b}.

\section{Preliminaries}
\label{sec:prelim}

\begin{figure}[h]
\centering
\includegraphics [width=5cm]{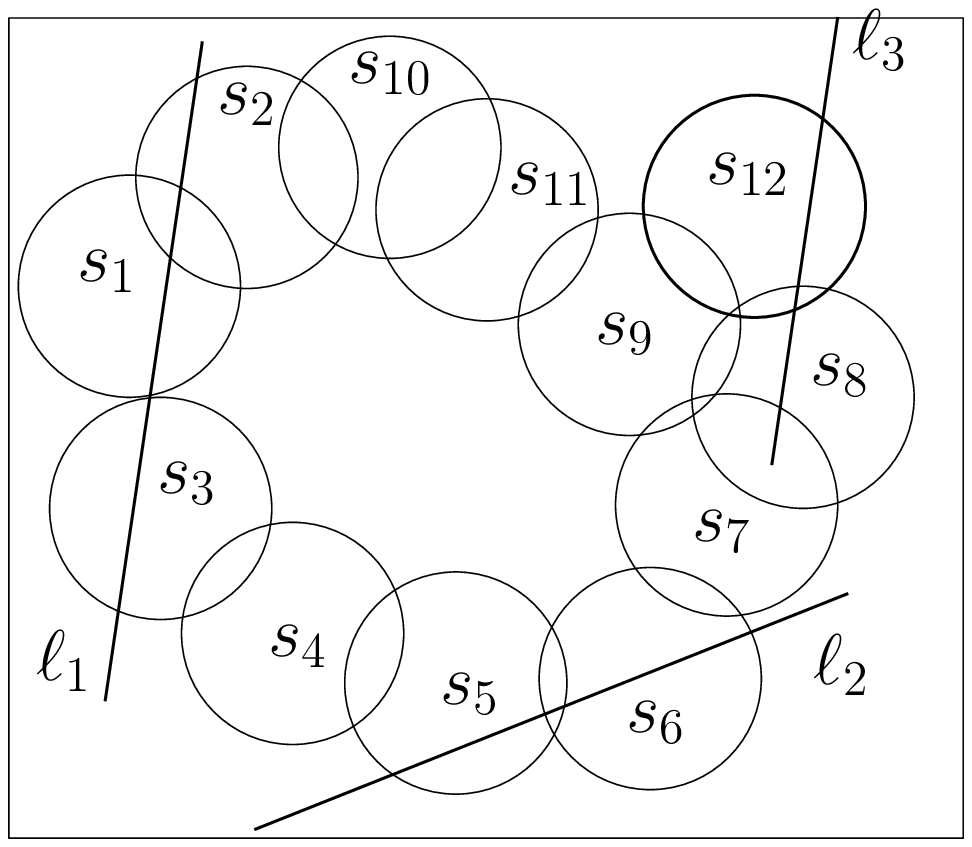} \hspace{0.5in} 
\includegraphics [width=3cm]{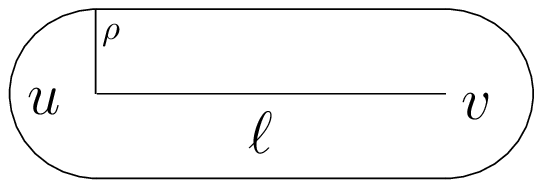}

\hspace{1cm} (a) \hspace{5cm} (b)
\caption{(a) Covering line segments by sensors, (b)
hippodrome of a line segment $\ell$} \label{lines_cover}
\vspace{-0.1in}
\end{figure}

We assume that the sensors are points in the plane. The sensing range of 
a sensor $s$ is a real number $\rho(s)$ (say), such that it can sense 
inside a circular region of radius $\rho(s)$. We assume that the sensing 
range of all the deployed sensors are the same, and is equal to $\rho$. 
A line $\ell$ is said to be covered by a sensor $s$ if there is at least 
one point on $\ell$ whose distance from $s$ is less than or equal to 
$\rho$. In other words, $\ell$ has intersection with the circle of radius $\rho$ 
centered at $s$. Figure \ref{lines_cover}(a) shows an example in which 
the lines  $\ell_1$ and $\ell_3$  are covered by 3 sensors while the line 
$\ell_2$ is covered by 2 sensors.

Given a line segment $\ell = [u,v]$ and a positive real $\rho$, the
hippodrome $H(\ell,\rho)$ is the union of all the points that are at
distance less than or equal to $\rho$ from some point in $\ell$ (see
Figure \ref{lines_cover}(b)). A line segment $\ell$ will be covered by
a sensor $s$ with sensing range $\rho$ if and only if $s$ is placed
inside the hippodrome $H(\ell,\rho)$. We need the following definition
for proving the NP-hardness result for the Line-Covering problem.

\begin{figure}[h]
\vspace{-0.2in}
\centering
\includegraphics [height=3.5cm]{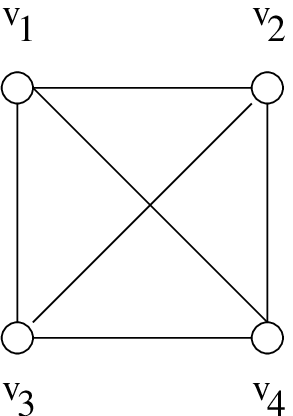} \hspace{2.5cm}
\includegraphics [height=3.5cm]{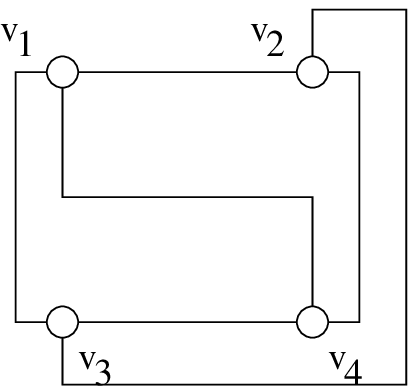}

(a) \hspace{5.5cm}  (b)

\vspace{0.5cm}
\includegraphics [width=4cm]{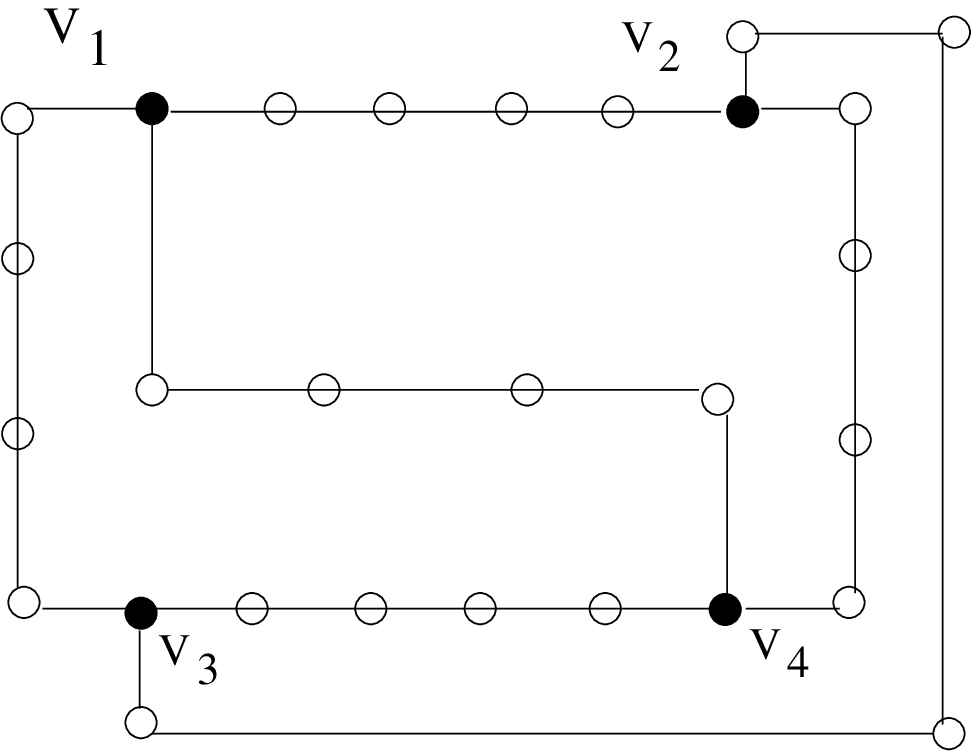} \hspace{1cm}
\includegraphics [width=4cm]{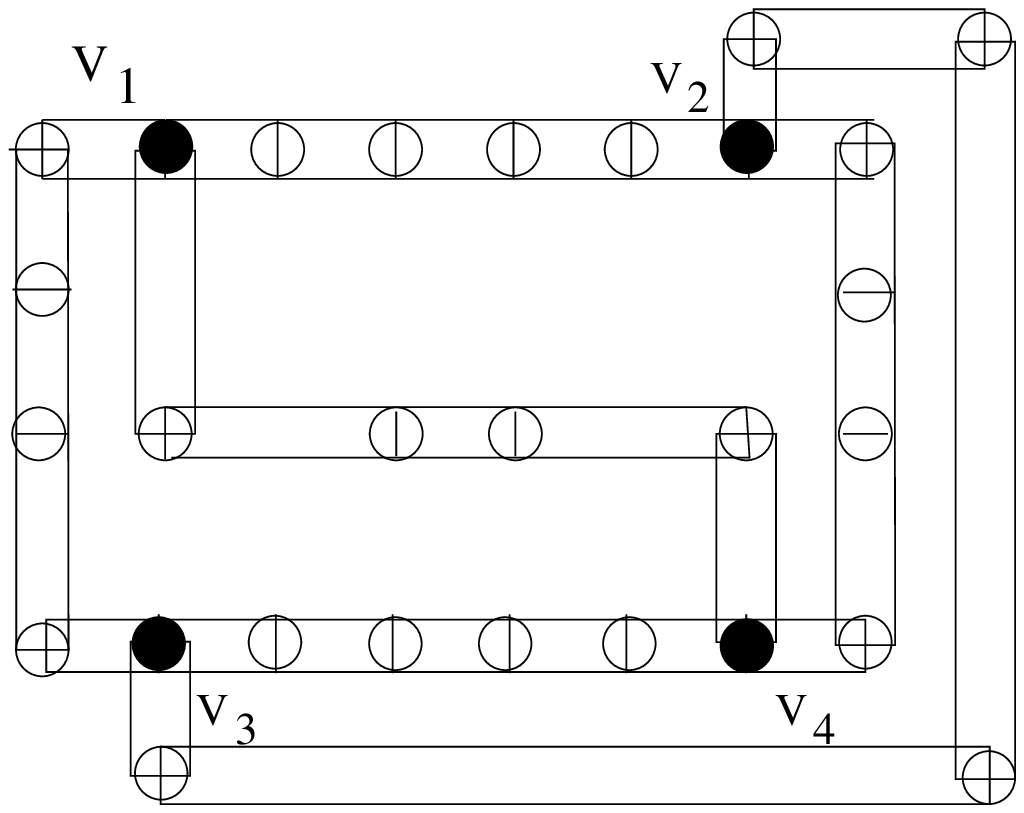}

(c) \hspace{5.5cm} (d)

\caption{(a) A cubic planar graph $G$, (b) 
its planar grid embedding $\mathcal{E}$, (c) its augmented planar grid
embedding $\mathcal{E}_{aug}$, and (d) hippodromes of
the edges in $\mathcal{E}_{aug}$} \label{cpg}
\vspace{-0.2in}
\end{figure}

\begin{definition}
A planar graph $G=(V,E)$ is said to be a \emph{cubic planar graph} if
the degree of each vertex $v \in V$ is at most three.
\end{definition}
\vspace{-0.1in}
A {\em planar grid embedding} of a graph is an embedding of the graph
in a grid such that the vertices of the graph are mapped to some grid
points, and the edges of the graph are mapped to non-intersecting grid
paths. Figure \ref{cpg}(a) shows a complete graph of four vertices,
and Figure \ref{cpg}(b) shows a planar grid embedding of it. In any 
planar grid embedding of a planar graph, one of the metrics of interest 
is the maximum number of bends along an edge in the embedding. The 
planar grid embedding of any cubic planar graph can be obtained in 
linear time \cite{tamassia1989}. Moreover, this algorithm ensures that 
the number of bends on each edge of the embedding is at most four. Thus, 
we have the following result:
\begin{result} 
The number of line
segments required to draw an edge in the planar grid embedding of a
cubic planar graph is at most five.
\end{result}

\vspace{-0.1in}
\section{Complexity results of Line-Covering problem}\vspace{-0.1in}
\label{sec:complexity}

In this section, we prove that the decision version of the {\em
Line-Covering} problem is NP-complete. We propose a polynomial time 
reduction from the vertex cover problem of a cubic planar graph 
to an instance of the {\em Line-Covering} problem. Needless to 
mention that the vertex cover problem for a cubic planar graph is 
NP-complete \cite{uehara1996}. We first give a polynomial-time 
reduction for obtaining an \emph{augmented planar grid embedding} 
of a cubic planar graph by using the planar grid embedding result 
of \cite{tamassia1989}. Next, we show that the original cubic planar 
graph $G=(V,E)$ has a vertex cover of size $\tau$ if and only if all 
the line segments in the embedding are covered by $\tau+2|E|$ sensors 
of a suitably chosen range $\rho$. Our proof is motivated by the work 
of Chabert and Lorca \cite{chabert2009}.

\vspace{-0.1in}
\subsection{ Polynomial time reduction }\label{subsec:ptrans}
\vspace{-0.1in}
Let $G=(V,E)$ be a connected cubic planar graph. We can generate a
planar grid embedding $\mathcal{E}$ of $G$ in linear time
\cite{tamassia1989}. Now, we execute the following two steps on 
$\mathcal{E}$ to obtain an {\em augmented planar grid embedding}
$\mathcal{E}_{aug}$ of the graph $G$.

\vspace{-0.1in}
\begin{description}
\item[Step 1:]
Add a new vertex at every bend of the embedding $\mathcal{E}$. Thus, 
each edge in the augmented embedding is either a horizontal or a
vertical line segment. These newly added vertices in
$\mathcal{E}_{aug}$ were not present in the vertex set $V$.
\item[Step 2:]
For every edge $(u,v) \in E$, identify the shortest path in
$\mathcal{E}_{aug}$ between $u$ and $v$. If the number of edges ($\alpha$)
in this path is less than five, then further augment
$\mathcal{E}_{aug}$ by adding $(5-\alpha)$ vertices on any edge of that
path to make the path length equal to five.
\end{description}
\vspace{-0.1in}

Figure \ref{cpg}(c) shows the augmented planar grid embedding of the
planar graph $G$ in Figure \ref{cpg}(a). Each node in $V$ is colored
with black, and each node added in augmentation steps 1 and 2 is
colored with white. Each black node has degree 3 and each white node
has degree 2. Each edge in $E$ corresponds to a chain of 5 edges in
$\mathcal{E}_{aug}$. Thus, the number of edges in $\mathcal{E}_{aug}$
is exactly $5|E|$. 

Observe that, the embedding $\mathcal{E}_{aug}$ contains both horizontal 
and vertical edges, and the grid size is polynomial in the number of 
vertices in the cubic planar graph $G$. Let $d$ be the length of the 
smallest edge in the embedding $\mathcal{E}_{aug}$. We choose a range 
$\rho < \frac{d}{2}$ for the covering problem. This ensures that 
the hippodromes $H(e,\rho)$ and $H(e',\rho)$ for a pair of edges $e,e' 
\in \mathcal{E}_{aug}$ do not intersect unless they share a common 
vertex (see Figure \ref{cpg}(d)). All the edges sharing a vertex $v 
\in \mathcal{E}_{aug}$ can be covered by placing a sensor anywhere 
in the intersection region of the hippodromes corresponding to these 
edges. Surely the vertex $v$ will lie in this region, and such a 
placement of sensor will be referred to as {\em placing a sensor at 
vertex $v$}. 

\vspace{-0.1in}
\begin{lemma}\label{lemma:NP-proof}
Given a positive integer $\tau \le |V|$, the planar graph $G$ has a
vertex cover of size $\tau$ if and only if the edges of the corresponding
$\mathcal{E}_{aug}$ can be {\em covered} using $(\tau+2|E|)$ sensors.
\end{lemma}
\vspace{-0.1in}

\begin{proof}
[$\Rightarrow$] Let $ \{v_1^\prime, v_2^\prime, \ldots,
v_\tau^\prime\}$ be a vertex cover of size $\tau$ in the graph $G$.
Deploy one sensor at the vertex corresponding to $v_i^\prime$ in
$\mathcal{E}_{aug}$, for each $i = 1, 2, \ldots, \tau$. Now, consider
any edge $e \in E$. Among the five edges in $\mathcal{E}_{aug}$
corresponding to the edge $e \in E$, at least one edge is already
covered by one of these $\tau$ sensors. To cover the remaining four
edges, only two sensors are sufficient, by placing one sensor in every
alternate vertex in the path. Hence, a total of $(\tau+2|E|)$ sensors
are sufficient to cover all line segments in  $\mathcal{E}$.

[$\Leftarrow$] Let there be a deployment of $(\tau+2|E|)$ sensors such
that each edge of $\mathcal{E}_{aug}$ is covered by at least one
sensor. Now, consider any edge $e = (u,v) \in E$, and its
corresponding 5-edge path $p(u,v) = u \rightarrow v_1 \rightarrow v_2
\rightarrow v_3 \rightarrow v_4 \rightarrow v$ in $\mathcal{E}_{aug}$.
To cover all the five edges, at least 2 sensors must be placed at two
of the four intermediate vertices in the path. Therefore, at least
$2|E|$ sensors are used to cover the intermediate three line segments
corresponding to $|E|$ edges in the cubic planar graph. Also, this
placement cannot cover both of the edges $(u, v_1)$ and $(v_4, v)$ of
$\mathcal{E}_{aug}$ at the same time, for each edge $(u,v) \in E$. By
the assumption, all these edges are also covered with $\tau$ sensors,
and surely these sensors are placed at {\it black} vertices. This
implies, we have a total of $\tau$ vertices in $G$ that covers all the
edges in $G$. \qed
\end{proof}

Given a deployment of sensors, verifying whether all lines are {\em 
covered} or not can be done in polynomial time. Thus, the covering 
decision problem for axis-parallel line segments is in NP. This leads 
to the following result: 

\vspace{-0.1in}
\begin{lemma}
Given a set of axis-parallel line segments, a real number $\rho$, 
an integer $\tau$, testing whether there exists a deployment of
$\tau$ sensors each of sensing range $\rho$, such that each member 
in $L$ is covered by at least one sensor, is NP-complete.
\end{lemma}

\vspace{-0.1in}
Since the problem of covering axis-parallel line segments is a special 
case of the {\em Line-Covering problem} for arbitrary line segments, the 
following theorem holds.
\begin{theorem}
The decision version of the {\em Line-Covering problem} is NP-complete.
\end{theorem}

\vspace{-0.1in}
\section{Approximation algorithm for covering axis-parallel
line segments}
\vspace{-0.1in}
\label{sec:approxalgo}
Here we present a 12-factor approximation algorithm for a special case of the 
{\em Line-Covering problem}, where the line segments inside the rectangular 
region $R$ are all axis-parallel. Following the method of \cite{ceriolia2004}, 
we first describe a 6-approximation algorithm for the case where the line 
segments are all horizontal. It is designed using the 2-factor approximation 
algorithm for covering the horizontal line segments in a strip of width 
$\sqrt{3}\rho$ as stated below. This concept is then extended to get the 
12-factor approximation algorithm when the line segments can be either 
horizontal or vertical.

\begin{figure}[t]
\vspace{-0.1in}
\centering
\includegraphics [width=4.5in]{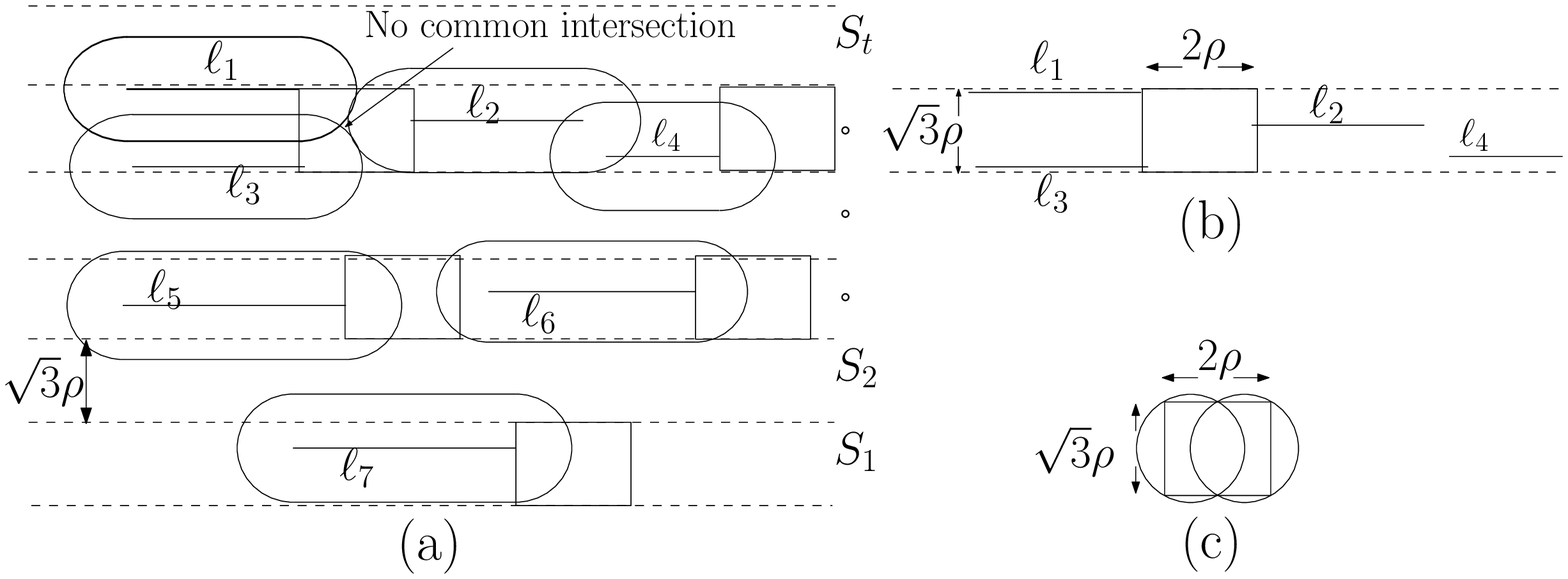} 

\caption{(a) Partitioning the area into horizontal strips, (b)
processing of $\ell_1$, (c) covering the line segments in $L_1$}
\label{coveringlines}
\vspace{-0.1in}
\end{figure}

We partition the entire region into horizontal strips $S_1, S_2, \ldots, S_t$, 
each of width $\sqrt{3}\rho$ (see Figure \ref{coveringlines}(a)). Thus, $t = 
\lceil\frac{h}{\sqrt{3}\rho}\rceil$, where $h$ is the height of the rectangular 
region $R$. Let $L_i=\{\ell_1 = (u_1,v_1),\ell_2 = (u_2, v_2), \ldots, \ell_r =
(u_r, v_r)\}$ be the line segments in a horizontal strip $S_i$, such that $v_1 
\leq v_2 \leq v_3 \leq \ldots \leq v_r$ (i.e., the line segments are sorted in 
increasing order of their right endpoints). We first choose $\ell_1$. Let $L_i^1$ 
denote the set of line segments in $S_i$ whose hippodromes intersect the hippodrome 
$H(\ell_1,\rho)$. In other words, each line segment $\ell \in L_i^1$ has at least 
one point which is at a distance at most $2\rho$ from $v_1$. Thus, all the line 
segments in $L_i^1$ intersect a rectangle $C$ of size $2\rho \times \sqrt{3}\rho$, 
inside the strip $S_i$ with left boundary at $v_1$ (see Figure \ref{coveringlines}(b)). 
If the hippodromes $H(\ell,\rho)$ for all the line segments $\ell \in L_i^1 \cup 
\{\ell_1\}$ share a common region, a sensor can be placed in that region to cover all 
the line segments in $L_i^1 \cup \{\ell_1\}$. However, such a favorable situation may 
not happen as shown in Figure \ref{coveringlines}(a). But, a rectangle of size $2 \rho 
\times \sqrt{3}\rho$ can be covered by only two circles of radius $\rho$ as shown in 
Figure \ref{coveringlines}(c). Thus, in order to cover all the line segments in $L_i^1 
\cup \{\ell_1\}$, one sensor is always necessary, and two sensors are always sufficient. 
We include the centers of those two circles in the set $Q_i$, the sensor positions for 
covering the horizontal line segments in the strip $S_i$. 

Next, we delete all the line segments in $L_i^1 \cup \{\ell_1\}$, and repeat the same 
steps with the remaining set of line segments in this strip. In each step, we add two 
sensor positions in $Q_i$. The process is repeated until all the members in $L_i$ are 
exhausted. If $Z_i$ is the number of iterations of the above steps required for 
processing the strip $S_i$, then $|Q_i| = 2Z_i$ is a loose upper bound and $Z_i$ is a 
loose lower bound on the number of sensors required to cover the line segments in the
strip $S_i$.

\begin{lemma} \label{extra} 
If $t$ denotes the number of horizontal strips of width $\sqrt{3}\rho$
in the region $R$, then 
\vspace{-0.1in}
\begin{itemize}
\item[(a)] $\cup_{j=1}^tQ_j$ is a 6-factor approximation solution for 
covering all the horizontal line segments in $R$, and
\item[(b)] $\frac{1}{3}\sum_{j=1}^tZ_j$ is a loose lower bound 
on the number of sensors required for covering all the horizontal line
segments in $R$.
\end{itemize}
\end{lemma}
\vspace{-0.1in}

\begin{proof}
Let $Q^h = \cup_{j=1}^t Q_j$ denote the set of sensors used to cover 
all the horizontal line segments by our algorithm. We use $Q_j'$ to 
denote the optimum set of sensors to cover the line segments in $L_j$ 
only, and $Q_j^*$ to denote the set of sensors in the optimum set of 
sensors for covering all the horizontal line segments in $R$, that 
are placed in the strip $S_j$. It is already mentioned that $|Q_j| 
\leq 2|Q_j'|$. Notice that, in order to cover the line segments in $L_j$ 
one needs to place sensors in $S_{j-1}$, $S_j$ and $S_{j+1}$. Thus, 
$|Q_j'| \leq |Q_{j-1}^*| + |Q_j^*| + |Q_{j+1}^*|$. Summing over all $j$, 
we have $\sum_{j=1}^t|Q_j'| \leq 3\sum_{j=1}^t|Q_j^*|$. Thus, 
$\sum_{j=1}^t |Q_j| \leq 6 \sum_{j=1}^t |Q_j^*|$.

The lower bound result follows from the facts that (i) a circle of 
radius $\rho$ centered at the center of a $2\rho \times \sqrt{3}\rho$ 
box in the strip $S_j$ does not cover any horizontal line segment in
$\cup_{i=1}^t L_i \setminus \{L_{j-1} \cup L_j \cup L_{j+1}\}$, and
(ii) $Z_j$ is the lower bound on the number of sensors required to
cover all the horizontal line segments in $S_j$. 
\qed
\end{proof}

The same technique can be adopted for covering all the vertical line
segments in the region $R$, and $Q^v$ is the set of sensor positions
for covering the vertical line segments in $R$, reported by our
algorithm. The following theorem summarizes the result in this
section.
\vspace{-0.05in}
\begin{theorem}
\label{th:12approx} $|Q^h\cup Q^v| \leq 12\times OPT$, where $OPT$
is the minimum number of sensors required to cover all the axis-parallel line
segments in $R$. The time complexity of our algorithm is $O(n\log
n)$, where $n$ is the number of line segments in $R$.
\end{theorem}
\vspace{-0.1in}
\begin{proof}
Let $Q^*$ denote the optimum set of sensors required to cover all 
the line segments in $R$, such that $|Q^*|=OPT$. Let $Q^{h*}$ and 
$Q^{v*}$ denote the set of sensors in the optimum solution $Q^*$ 
such that all the horizontal line segments are covered by the sensors 
in $Q^{h*}$, and all the vertical line segments are covered by the 
sensors in $Q^{v*}$. Thus, $Q^* = Q^{h*} \cup Q^{v*}$, where $Q^{h*}
\cap Q^{v*}$ may not be empty. Since $|Q^*| \ge max(|Q^{h*}|,|Q^{v*}|)$ 
and $|Q^h\cup Q^v| \le (|Q^h| + |Q^v|)$, therefore $|Q^h\cup Q^v| \le 
2*max(|Q^h|,|Q^v|) = 2*max(6|Q^{h*}|,6|Q^{v*}|) \leq 12|Q^*|$. 

The time complexity result follows from the following two facts: (i) placing
each horizontal (resp. vertical) line segment in appropriate strip
needs $O(n_h)$ (resp. $O(n_v)$) time, where $n_h$ (resp. $n_v$) is the
number of horizontal (resp. vertical) line segments in $R$, and (ii)
processing the line segments in a horizontal (resp. vertical) strip
takes $O(m\log m)$ time, where $m$ is the number of line segments in
that strip. \qed
\end{proof}

\vspace{-0.1in}

\section{PTAS for covering axis parallel line segments }
\vspace{-0.1in}
Let $H$ denote the set of hippodromes corresponding to all the
horizontal and vertical line segments in $R$. Let $X=\{x_0,x_1,x_2,
\ldots x_m\}$ be a sorted sequence of distinct real numbers, where
$x_0$ and $x_m$ correspond to the $x$-coordinates of the left and
right boundaries of the region $R$, and $x_i, i=1, 2,\ldots, m-1$  
denote the $x$-coordinates of the corners of the bounding rectangles
of these hippodromes. We need to find a minimum size set $P$ of
points, called the minimum piercing set for $H$, such that each
hippodrome in $H$ contains at least one point in $P$. In other words,
if we position one sensor in each point of $P$, each line segment in
the region $R$ will be covered by at least one sensor. 

We show the existence of a PTAS for problem of covering  
axis-parallel line segments by (i) proposing an algorithm for 
computing a piercing set for $H$ of size $(1+\epsilon)\times OPT$, 
where $OPT$ is the size of the optimum piercing set for $H$, and 
$\epsilon$ is a desired small real number, and (ii) proving the 
time complexity of the algorithm to be a polynomial function of $n$, 
where the degree of the polynomial may depend on $\epsilon$, 
assuming that the sensing range $\rho$ is not too small compared 
to the floor dimensions. In particular, we assume that $\frac{h}{\rho}$ 
is a constant, where $h$ is the height of $R$. Our algorithm uses 
the idea from \cite{chan2005}.

\begin{figure}[h]
\centering
\includegraphics [width=4.5in]{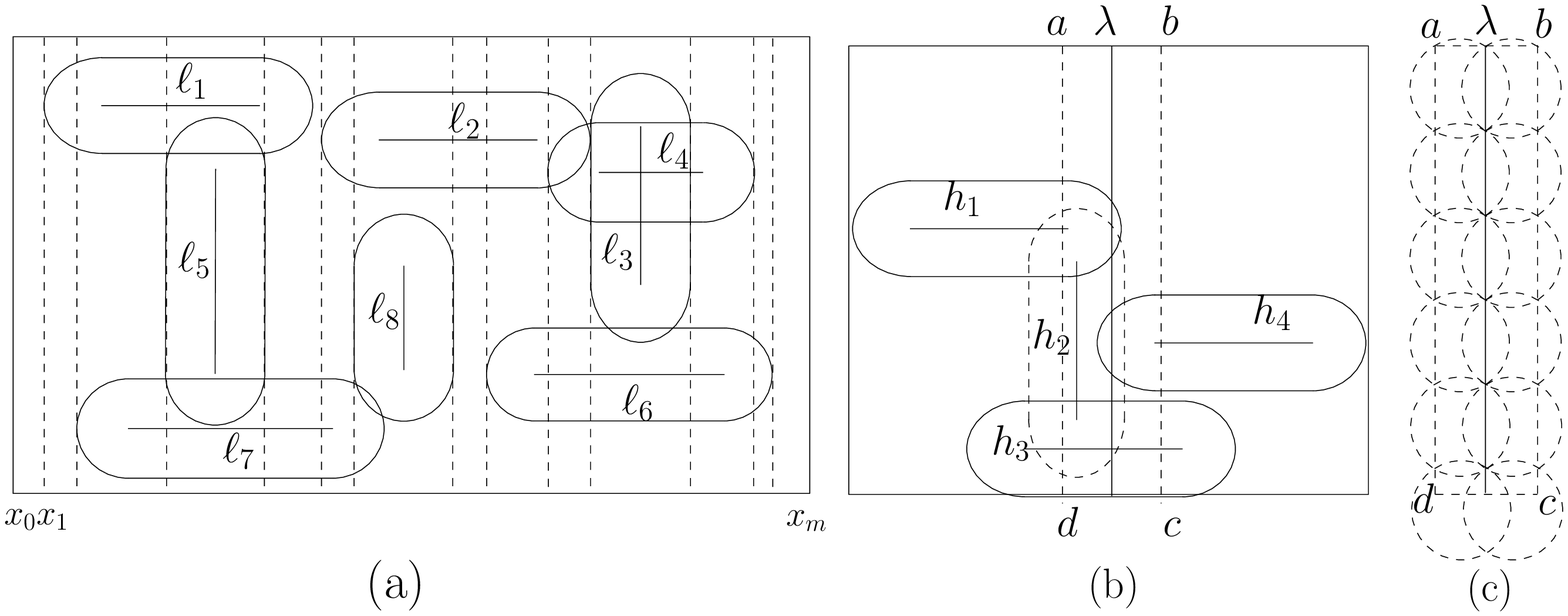} 

\caption{(a) Splitting of $R$ into vertical strips, 
(b) set of hippodromes intersected by a vertical line, 
and  (c) demonstration of covering the hippodromes 
intersected by a vertical line by $2t$ sensors}
\label{fig:ln_hippo_intr}
\vspace{-0.1in}
\end{figure}

\begin{lemma} \label{lx}
If the region $R$ has $t$ number of horizontal strips of width 
$\sqrt{3}\rho$, then 
a set of $2t$ points always pierce all the
hippodromes that are intersected by any vertical line inside $R$.  
\end{lemma}
\vspace{-0.1in}
\begin{proof}
In Figure \ref{fig:ln_hippo_intr}(b), a situation is depicted where 
a set of hippodromes are intersected by a vertical line $\lambda$. 
All these hippodromes can be pierced by placing the points at the 
centers of $t$ pairs of circles (of radius $\rho$) shown in Figure 
\ref{fig:ln_hippo_intr}(c), where $t$ is the number of horizontal
strips of width $\sqrt{3}\rho$. Since $h$ is the height of $R$, $t =
\frac{h}{\sqrt{3}\rho}$ \qed 
\end{proof}

\vspace{-0.1in}
\subsection{Algorithm}
\vspace{-0.1in}
We partition the region $R$ by drawing vertical lines at 
$x_0,x_1,x_2,\ldots,x_m$ (see Figure \ref{fig:ln_hippo_intr}(a)). 
Let $P$ be a set of points that accumulates the piercing points 
obtained by this algorithm. We fix a vertical line $\ell$ at $x_0$. 

For each $i=1,2,\ldots m$, we execute the following two steps: 
\begin{enumerate}
\item[Step 1:] Let $\Theta$ denote the subset of hippodromes that are 
intersected by the vertical line $x=x_i$, and \\
$\Theta'$ be the set of hippodromes which are properly inside the  
strip defined by the vertical lines $x=\ell$ and $x=x_i$. None of these
hippodromes intersect the vertical lines $x=\ell$ and $x=x_i$
\item[Step 2:] Compute the lower bound $\psi$ on the number of sensors required
to pierce all the hippodromes in $\Theta'$ using Lemma \ref{extra}(b).
If $\psi$ is greater than or equal to some constant $T$ 
(decided a priori) or $i=m$, then

\begin{enumerate}
\item put $i$ in an array $I$. 
\item Get a set $\Pi$ of points to pierce the hippodromes in
$\Theta$ using Lemma \ref{lx}.
\item Compute the exact minimum piercing set $\Pi'$ of $\Theta'$ by
an exhaustive search algorithm discussed in the proof of Lemma
\ref{ly}.
\item Set $P=P \cup \Pi \cup \Pi'$, and $\ell=x_i$.
\end{enumerate}
\end{enumerate}

Thus, we have computed the optimal piercing set for the hippodromes
in each vertical strip defined by the vertical lines $x=x_{I[j-1]}$
and $x=x_{I[j]}$, for $j = 1, 2, \ldots |I|$.

\begin{lemma} \label{ly}
If $n_j=|\Theta_j'|$ for the vertical strip $V_j$ bounded by two
vertical lines $x=x_{I[j-1]}$ and $x=x_{I[j]}$, then the time
complexity of finding the minimum piercing set (of points) for the set
of hippodromes $\Theta_j'$ is $O(n_j^{24T+4t-1})$, where
$t=\frac{h}{\sqrt{3}\rho}$.
\end{lemma}
\begin{proof}
Let $Q$ be the set of intersection points of the hippodromes in
$\Theta_j'$. Since $n_j = |\Theta_j'|$, $|Q| = O(n_j^2)$ 
in the worst case, and by a brute-force method, these can be found 
in $O(n_j^2)$ time. These points may be used as piercing points.

Let us denote by $P_{\alpha,\beta}'$ the smallest piercing set of the
set of hippodromes that lie properly inside $x=x_\alpha$ and
$x=x_\beta$, and $P_\alpha$ denote the smallest piercing set of the
set of hippodromes intersected by the vertical line $x=x_\alpha$.

Let us consider the vertical strip $V_j$ defined by the
vertical lines $x=x_\alpha$ and $x=x_\beta$, where $\alpha =
I[j-1]$ and $\beta=I[j]$. Let $\gamma=\beta-1$. 

Surely, $|P_{\alpha,\beta}'| \leq |P_{\alpha,\gamma}'| + |P_\gamma| \leq
|P_{\alpha,\gamma}'| + 2t$ (by Lemma \ref{lx}). If our approximation 
algorithm returns a lower bound $T$ for the size of the solution for 
the set of hippodromes $\Theta_j'$, we have $|P_{\alpha,\gamma}'| \leq
12T-1$. Thus the number of points in the optimum piercing set for
$\Theta_j'$ is $|P_{\alpha,\beta}'| \leq 12T-1+ 2t$. So, we need to 
consider all possible subsets of size less than or equal to
$|P_{\alpha,\beta}'|$ (= $\mu$ say) from $Q$. The number of such
subsets is ${|Q|} \choose \mu$ = $O(n_i^{2\mu})$ in the
worst case. Since, for each subset, we need to check whether it pierces
all the hippodromes in $\Theta_j'$, the overall complexity of getting
the smallest piercing set of $H_i'$ is $O(n_i^{2\mu+1})$, where $\mu
\leq 12T-1+ 2t$. Thus the lemma follows. \qed
\end{proof}

We now show that for a given set of $n$ axis-parallel line segments,
the algorithm {\em COVER-PTAS} produces a $(1+\epsilon)$-factor
approximation result. 

Observe that, we have computed the optimum cover solution in each
vertical strip defined by the elements in array $I$. We have also
shown that the hippodromes intersecting the vertical lines
$x=x_{I(i)}, i \in I$ can be covered by a set of $2t\times (|I|-1)$
sensors, where $t=\frac{h}{\sqrt{3}\rho}$. Actually, our algorithm
{\em COVER-PTAS} produces a solution of size $OPT' =
\sum_{i=1}^{|I|-1} |P_{i,i+1}'| + 2t\times (|I|-1)$. If
$OPT$ is the minimum number of sensors required to cover the line
segments in $L$, then $$\sum_{i=0}^{|I|-1} |P_{i,i+1}'| \leq
OPT \leq \sum_{i=0}^{|I|-1} |P_{i,i+1}'| + 2t\times (|I|-1)
= OPT'.$$

By Step 3(b) of our algorithm $|P_{i,i+1}'| \geq T$. Since
the size of the optimum solution is $OPT$, the number of disjoint
strips $|I|-1 \leq \frac{|OPT|}{T}$.  

Thus, $|OPT'| \leq  |OPT| + 2t \times (\frac{|OPT|}{T})$
= $|OPT| \times (1+\frac{2t}{T})$. For a given $\epsilon$, we may get an 
$(1+\epsilon)$-factor approximation result by choosing $T
= \frac{2t}{\epsilon}$.

In order to compute the array $I$, we have to execute the 12-factor approximation 
algorithm $2n$ times. This needs $O(n^2\log n)$ time in the worst case. 
The time complexity for computing $P_{i,i+1}'$ for all $i =
1, 2, \ldots, |I|-1$ needs $O(\sum_{i=0}^{|I|-1} n_i^{24T+4t-1})$,
which may be $O(n^{24T+4t-1})$ in the worst case. Thus, we have the 
following result:
\begin{theorem}
\label{th:ptas} Given a set of $n$ axis-parallel line segments, the algorithm 
{\em COVER-PTAS} produces a solution (placement of sensors for covering 
all the line segments) of size $(1+\epsilon)OPT$ in time $O(n^2\log
n + n^{(\frac{24}{\epsilon}T + 4t - 1)})$, where $OPT$ is the size of
the optimum solution, and $t = \frac{h}{\sqrt{3}\rho}$ = the number of
horizontal strips required to partition the region $R$ of width
$\sqrt{3}\rho$, and $T=\frac{2t}{\epsilon}$.
\end{theorem}

\section{PTAS for 1-line covering line segments of any arbitrary orientation}

There is a given a set of line segments $L=(l_0,l_1, \ldots l_n)$ in a bounded rectangular region $R$. The line segments 
are arbitrarily oriented and whose lengths are at most some constant $c$ times of the sensors' sensing range $s_r$ 
(i.e $ length(l_i) \le c \times s_r$ for $i=1$ to $n$). 
Let $H$ denote the set of hippodromes corresponding to the line segments. We need to find a minimum size set $P$ of points, called the minimum 
piercing set for $H$, such that each hippodrome in $H$ contains at least 
one point in $P$. In other words, if we position one sensor in each point of $P$, each line segment in the region $R$ will be covered by at 
least one sensor. We now propose an algorithm for computing a piercing set for $H$, and show that it produces a solution of size 
$(1+\frac{1}{k})\times |OPT|$, where $OPT$ is the minimum piercing set for $H$, and $k$ is a desired integer number. 

\begin{figure}[h]
\centering
\includegraphics[width=12cm]{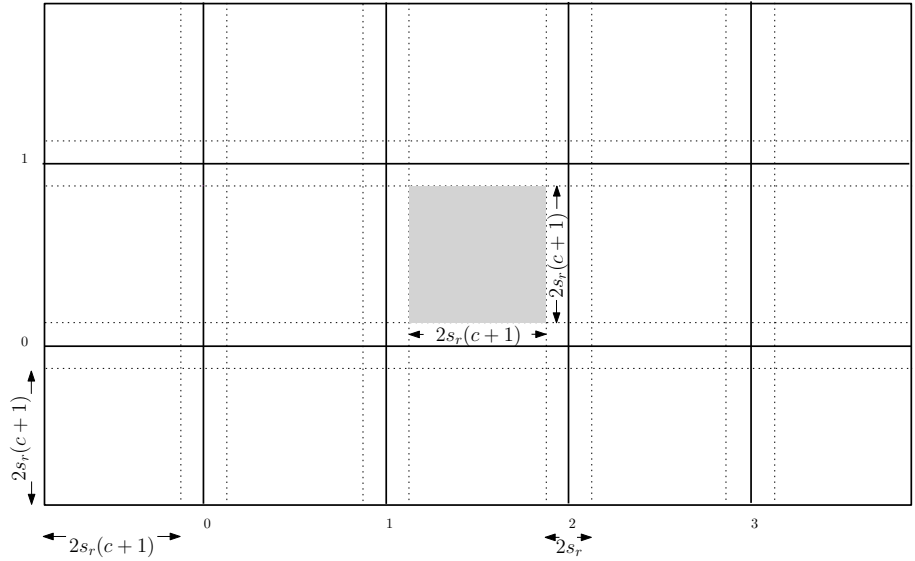} 
\caption{Splitting of $R$ into horizontal and vertical strips }
\label{fig:arbitrary_ln}
\end{figure}

{\bf Algorithm 1-COVER-PTAS}

\begin {enumerate}

\item Initialize the piercing set $P=\emptyset$.

\item Partition the region $R$ by drawing horizontal and vertical strips of width $2s_r$ in the interval of $2s_r(c+1)$. 
Let the region $R$ be subdivided into $M \times N$ square blocks after placing the strips. (see Figure \ref{fig:arbitrary_ln}).
The horizontal strips are numbered from bottom to top  $(0,1,\ldots M-1)$ and the vertical strips are numbered from left to right
 $(0,1,\ldots N-1)$.

\item Group the set of the vertical strips into $k (\ge 1)$ disjoint subgroups $(VST_0,$
$VST_1, \ldots VST_{k-1})$. 
Subgroup $VST_i$ contains vertical strips whose indices are $(i,i+k,i+2k, \ldots )$.

\item Let $s_i$ denote the optimum number of sensors required to cover the line segments that intersect vertical strip $i$.  

\item The set of line segments that intersect the vertical strips are also partitioned into $k$ disjoint subsets 
$(LS_0, LS_1, \ldots LS_{k-1})$. All the line segments in $LS_i$ must intersect one and only one vertical strip in $VST_i$.
Hence, the line segments in $LS_i$ can be covered optimally using $OPT_i = (s_i \cup s_{i+k} \cup \ldots )$ sensors.

\item Select a $t$ among $o$ to $k-1$ for which $|OPT_t|$ is minimum.

\item Set $P=OPT_t$.

\item Remove the set of line segments in $LS_t$.

\item The region $R$ is subdivided into small subregions $(R_1,R_2, \ldots )$ by the vertical strips in $VST_t$. 

\item The remaining line segments $L \setminus LS_t$ are  fully inside these subregions.

\item For each individual disjoint subregion $R_i$.

\begin{enumerate}
  \item Let $\overline{OPT_i}$ denote the minimum number of sensors required to cover the line segments inside $R_i$.
  \item $P= P \cup \overline{OPT_i}$.
\end{enumerate}
\end{enumerate}

\begin{lemma} 
\label{lem:lx1}
Let $OPT$ denote the minimum-size set of sensors which covers all the line segments in $L$ then there exists a subgroup $LS_t$ which can be 
covered using at most $\frac{|OPT|}{k}$ sensors. 
\end{lemma}
\begin{proof}
Since $OPT$ covers all the line segments in $L$; therefore, it also covers all the line segments in $L_{vs}=(LS_0 \cup LS_1 \cup \ldots LS_{k-1})$. 
As the hippodromes corresponding to the line segments between any two subset $LS_i$ and $LS_j$ are disjoint where $i$ and $j$ are in between 
$i=0$ to $(k-1)$  and $i \ne j$. Therefore, $(|OPT_0| + |OPT_1| + \ldots |OPT_{k-1}|) \le |OPT|$. Hence, the minimum among $(|OPT_0|,|OPT_1|, \ldots |OPT_{k-1}|)$ 
must be at most $\frac{|OPT|}{k}$.
\end{proof}

\begin{theorem}
\label{the:th1}
The size of the piercing set $P$ returned by our algorithm is at most $(1+\frac{1}{k})|OPT|$.
\end{theorem}
\begin{proof}
After execution of step(9) of our algorithm the whole region is partitioned into disjoint subregions $(R_1, R_2, \ldots )$. 
Let $(\overline{OPT_0},\overline{OPT_1}, \ldots)$ denote the optimum number of sensors used to cover the line segments which are fully inside each of the 
individual subregions. The subregions are also separated by the distance $2s_r$. Therefore, $(|\overline{OPT_0}| + |\overline{OPT_1}| + \ldots) \le |OPT| $. 
The number of sensors used to cover the line segments that intersect the vertical strips in $LS_t$ is at most $\frac{|OPT|}{k}$. Hence, the overall number 
of sensors used by our algorithm is at most $|OPT|+\frac{|OPT|}{k}= (1+\frac{1}{k})|OPT|$.  
\end{proof}

\begin{lemma} 
\label{lem:lx2}
If $B$ is a square region of size $2s_r(c+1) \times 2s_r(c+1)$ then the number of sensors required to cover all the line segments 
that are totally inside $B$ is at most $2(c+1)^2$. Similarly, the set of line segments that intersect a rectangular  strip of size $2s_r(c+2) \times 2s_r$ 
can be covered by covering the rectangular strip and it requires at most $\frac{4}{\sqrt{3}}(c+2)$ sensors.   
\end{lemma}
\begin{proof}
The maximum size of the square that is inside a circle of radius $s_r$ is $\sqrt{2}s_r \times \sqrt{2}s_r$. Therefore, the number of sensors
required to fully cover $B$ is $\frac{4s_r^2(c+1)^2}{2s_r^2}= 2(c+1)^2$. There is a rectangle of size $\sqrt{3} s_r \times s_r$
which is totally inside a circle of radius $s_r$. Therefore, the number of sensors used to cover the whole strip is 
$\frac{2s_r(c+2)\times 2s_r}{\sqrt{3}s_r^2} = \frac{4}{\sqrt{3}}(c+2)$. 
\end{proof}

\begin{lemma}
\label{lem:lx3}
The time needed to cover $n_i$ line segments inside subregion $R_i$ is $O(n_i^{\frac{4}{\sqrt{3}}(2Mk-M-k)(c+2) + 2Mk(c+1)^2})$. 
\end{lemma}
\begin{proof}
Let $h$ denote the height of the rectangular region $R$ then the number of rows $M$ of square boxes is $\frac{h}{2s_r(c+2)}$. 
Therefore, the number of boxes in the subregion $R_i$ is at most $k*M$. We determine the optimum number of sensors needed to cover 
all the line segments in each individual box $B_{i,j}$. Let $m_{i,j}$ denote the number 
of line segments in a box $B_{i,j}$ then the time required to evaluate the optimum number of sensors needed to cover the line segments 
in $B_{i,j}$ is $O(m_{i,j}^{4(c+1)^2})$. Similarly, to cover a set of line segments $T_i$ which intersect a vertical strip of size 
$2s_r(c+2) \times 2s_r$ can be done in time $O(T_i^{\frac{8}{\sqrt{3}}(c+2)})$. \\
The overall sensors used to cover all the line segments inside $R$ is divided into two sub-parts. \\
$(i)$ The sensors used to cover line segments optimally inside individual box $B_{i,j}$ where $i \in [0,M-1]$ and $j \in [0,N-1]$. \\
$(ii)$ The sensors used to cover line segments optimally that intersects the intermediate strips. \\
In subregion $R_i$, there are at most $(k-1)$ vertical strips and $(M-1)$ horizontal strips. Therefore, number of strips of length $2s_r(c+2)$ 
in a subregion $R_i$ is at most $(2Mk-M-k)$ and number of boxes in a subregion 
is at most $Mk$. Hence, the upper bound on the number of sensor needed is $T=\frac{4}{\sqrt{3}}(2Mk-M-k)(c+2)+2Mk(c+1)^2$. If there are $n_i$ line 
segments in region $R_i$ then the time required to find the optimum number of sensors required to cover the line segments in $R_i$ is $n_i^T$. 
Therefore, the overall time complexity is $O(n_i^{\frac{4}{\sqrt{3}}(2Mk-M-k)(c+2) + 2Mk(c+1)^2})$.  
\end{proof}

\begin{lemma}
 The time needed to cover $m_i$ line segments that intersect a vertical strips in $VST_t$ is $O(m_i^{\frac{4}{\sqrt{3}}M(c+2)})$.
\end{lemma}
\begin{proof}
A vertical strip of height $h$ consists of $M$ strips each of height $2s_r(c+2)$. Therefore, the upper bound on the number of sensors needed
by Lemma \ref{lem:lx2} is $U=M*\frac{4}{\sqrt{3}}(c+2)$. 
So, by exhaustive search the time required to cover $m_i$ line segments is $O(m_i^{\frac{4}{\sqrt{3}}M(c+2)})$.
\end{proof}

\begin{theorem}
 The overall run time of the algorithm is $O(n^{\frac{4}{\sqrt{3}}(2Mk-M-k)(c+2) + 2Mk(c+1)^2} + n^{\frac{4}{\sqrt{3}}M(c+2)})$.
\end{theorem}
\begin{proof}
It is comprised of two times $(i)$ The time needed to cover individual subregions optimally and $(ii)$ The time needed to cover the line segments that 
intersects the vertical strips in $VST_t$ optimally. Hence, the total time is
$(n_0^T + n_1^T + \ldots) + (m_0^U + m_1^U + \ldots) \le n^T+n^U $
$= O(n^{\frac{4}{\sqrt{3}}(2Mk-M-k)(c+2) + 2Mk(c+1)^2} + n^{\frac{4}{\sqrt{3}}M(c+2)})$.

\end{proof}

\section{Discussion and Conclusion}
\label{sec:conclude}
\vspace{-0.1in}
The problem of covering a set of line segments with minimum number of 
sensors is introduced in this paper. The problem is proved to be NP-hard. 
We provided a 12-factor approximation algorithm for the special case where 
the line segments are axis-parallel. We have also shown that  
a PTAS exists for this problem. These problems are useful in the context 
of intruder detection in a restricted area. Developing an efficient 
algorithm with a good approximation factor for the general problem, where 
the line segments are of arbitrary orientation, and each line segment is 
covered by at least $k$ sensors. We have also shown a PTAS algorithm for 
aribitrary oriented line segments for segments whose length are at most some constant times of 
the sensing range. We would also want to improve the approximation 
factor of this restricted case (for axis parallel line segments and $k=1$). 

\bibliographystyle{abbrv}
\bibliography{linecover}  
\end{document}